\documentclass[conference]{IEEEtran}
\IEEEoverridecommandlockouts
\usepackage{cite}
\usepackage{amsmath,amssymb,amsfonts}
\usepackage{algorithmic}
\usepackage{graphicx}
\usepackage{textcomp}
\usepackage{subfigure}
\def\BibTeX{{\rm B\kern-.05em{\sc i\kern-.025em b}\kern-.08em
    T\kern-.1667em\lower.7ex\hbox{E}\kern-.125emX}}
\usepackage{url}
\usepackage{booktabs} 
\usepackage{colortbl}
\usepackage[ruled, vlined, linesnumbered]{algorithm2e}

\usepackage{multirow}
\usepackage{amsmath}
\usepackage{comment}
\usepackage{mathtools}
\usepackage{multirow}

\usepackage{colortbl}
\usepackage[dvipsnames]{xcolor}
\usepackage{balance}
\usepackage[top=0.8in, left=0.9in, right=0.9in]{geometry}

\begin{document}

\title{Context-aware Human Intent Inference for Improving Human Machine Cooperation }
\author{\IEEEauthorblockN{Xiang Zhang}
\IEEEauthorblockA{
\textit{School of Computer Science \& Engineering,}
\textit{University of New South Wales, Sydney, Australia} \\
xiang.zhang3@student.unsw.edu.au}
}

\maketitle




\begin{abstract}
The ability of human beings to precisely recognize others’ intents is a significant mental activity in reasoning about actions, such as, what other people are doing and what they will do next. Recent research has revealed that human intents could be inferred by measuring human cognitive activities through heterogeneous body and brain sensors (e.g., sensors for detecting physiological signals like ECG, brain signals like EEG and IMU sensors like accelerometers and gyros etc.). In this proposal, we aim at developing a computational framework for enabling reliable and precise real-time human intent recognition by measuring human cognitive and physiological activities through the heterogeneous body and brain sensors for improving human machine interactions,  and serving intent-based human activity prediction.



\end{abstract}

\section{MOTIVATION} 
\label{sec:introduction}

To reliably and precisely recognize human intents, we propose to develop a weakly supervised attention-based deep multi-task neural network, by jointly modelling shared commonality, specific disparity of multiple sources (e.g., brain activities, emotions, gaze tracking, head tracking, hand motion, head-pose estimation and pointing gesture recognition etc.), as well as inter-relations of across contexts (e.g., user physiological and mental status like stressful, locations and behaviours of daily routines etc.).
Due to the complexity of human-machine interactions and high uncertainties of capturing human intents, constructing a reliable real-time intent inference system that continuously interacts with an ever-changing environment, can be challenging. We identify the following four major challenges:

1) How to handle low signal-to-noise, interference and numerical drifts of heterogeneous sensors. 
Many existing works focus on developing careful signal processing design or sophisticated feature engineering; however, they might not be sufficient in some cases [2][3]. Incorporating additional source information (e.g., onset contexts) would be expected to handle the obstacle more adaptively and effectively; 

2) How to deal with the intra-class variability of intents? Intents may be varied among different individuals with distinct patterns [4][7]. 

3)    How to effectively fuse and distill heterogeneous sensor data to extract the relevant information with acceptable quality and integrity. 
When context changes, multi-sensor fusion methods need to deal with such changes as they can greatly affect the properties of the methods such as accuracy [5][6]. For example, correlations among sources are critical to make an appropriate judgment on intent detection, while may not be obvious and explicit; 


\section{PROPOSED APPROACH}
To address the identified challenges above and achieve the goal of this proposal, we conscientiously decompose it into three successive stages:


\subsection{Task 1. Multi-Task Intent Modelling}
The task is aimed at modeling intents from multi-source data. Information scattered in various sensors (e.g, physiological signals like ECG, brain signals like EEG and IMU sensors like accelerometers and gyros etc.) in fact describes the inherent characteristics of the same operator in various aspects. The motivation behind this task lies in that intent patterns are unveiled in the different domain, and these patterns may be complementary or mutually correlated. 
Hence, a person's intent should show source disparity, as well as consistency. To discover such invariant commonalities, we propose a generative adversarial network under the multi-task framework, wherein the generative inference model is focusing on predicting intent-specific patterns given the heterogeneous input, and the discriminative retrieval focusing on predicting source-specific intent patterns given the heterogeneous input. A game theoretical minimax process is used to iteratively optimize both of models jointly in an adversarial manner with adding orthogonality constraint. 

\subsection{Task 2.  Fusing Multi-Faceted Contextual Information}
To further enhance the intent inference accuracy, we propose to take into account the person’s location and historical behaviors to achieve improved results in terms of both inference accuracy and speed based on Task 1. The motivation behind this task is that human intent is tightly bound to contexts, interplayed and coexist with each other in terms of temporal and spatial dimensions, which provides a valuable clue to improve intent recognition in an inference logic with appropriate Bayesian fusion. 
We propose a hybrid scalable model by combining dynamic Bayesian network with Granger constraints and recurrent neural network to model diverse contextual dependencies. The proposed model is able to fuse the user status, context, and user's behavior intention, by explicitly allowing different types of links, representing different types of relations between various contexts, such as co-occurrence relationships among contexts.

\subsection{Task 3.  Calibrating via Ontological Reasoning for Reliable Intent Inference}
This task is towards the exploration of using of ontological reasoning to reinforce the performance of learning-based approaches. The intuition behind this idea is that certain human intent and behaviors can be inferred with his contexts without training.
To enhance algorithmic discovery techniques proposed in previous phases, we will construct a behavior intent knowledge base. One of the most important aspects of such a knowledge base is that it is structured around a semantic ontology, by organizing intents according to interactions where they usually take place. The ontology provides a rich hierarchy with multi-levels of depth in the context of human behavior intents in the specific domain. We will represent the domain in a set of first-order probabilistic logics, which can be further specified as the discriminative structure and qualitative parameter constraints on the intent model. We propose an ontological reasoning approach to effectively represent a context change or its situational variations. 

\section{Current Progress}
In my first year, I mainly focused on the EEG-based intent modeling.
We [8] develop a deep learning algorithm to directly works on the raw EEG data and obtain the classification accuracy around $95\%$ on the 5-class classification problem. Furthermore, as part of our future work [1], we design a unified deep learning framework that leverages the Recurrent Convolutional Neural Network (R-CNN) to capture spatial dependencies of raw EEG signals based on features
extracted by convolutional operations and temporal
correlations through RNN architecture, respectively.
The experimental
results illustrate that the proposed framework achieves highlevel
of accuracy over both the public dataset (95.53\%)
and the locally collected dataset (94.27\%).
In addition, the common intent modeling is a key technology on the practical intent-based human machine interaction. We [9] investigated to extract the common intent patterns of different subjects and enhance the intent recognition precision cross different individuals for better generalization and practical use in the real-world deployment.

\section{CONCLUSION}
This proposal is the very first approach that is presented as a systematic way of deeply integrating knowledge- and data-driven framework and set of techniques for human intent recognition over ambient and federated data sources. The proposed framework and techniques will provide foundations for transforming continuous intent inference from a process that is costly, complex and mostly expert-driven to one that is simple, rigorous, reusable, and flexible. 

\section{ACKNOWLEDGE}
Many thanks to my supervisor Dr. Lina Yao for her support and valuable guidance.

\section{REFERENCE}
[1] Zhang, X., Yao, L., Sheng, Q. Z., Kanhere, S.S, Gu, T. and Zhang, D., 2018, “Converting your thoughts to texts: Enabling brain typing via deep feature learning of EEG signals,” in IEEE International Conference on Pervasive Computing and Communications.

[2] Rodríguez-Bermúdez, G. and Garcia-Laencina, P.J., 2015. Analysis of EEG signals using nonlinear dynamics and chaos: a review. Applied Mathematics \& Information Sciences, 9(5), p.2309.

[3] Movassaghi, S., Abolhasan, M., Lipman, J., Smith, D. and Jamalipour, A., 2014. Wireless body area networks: A survey. IEEE Communications Surveys \& Tutorials, 16(3), pp.1658-1686.

[4] Yao, L., Nie, F., Sheng, Q.Z., Gu, T., Li, X. and Wang, S., 2016, September. Learning from less for better: semi-supervised activity recognition via shared structure discovery. In Proceedings of the 2016 ACM International Joint Conference on Pervasive and Ubiquitous Computing (pp. 13-24). ACM.

[5] Gravina, R., Alinia, P., Ghasemzadeh, H. and Fortino, G., 2017. Multi-sensor fusion in body sensor networks: State-of-the-art and research challenges. Information Fusion, 35, pp.68-80.

[6] Goodfellow, I., 2016. NIPS 2016 tutorial: Generative adversarial networks. arXiv:1701.00160.

[7] Bulling, A., Blanke, U. and Schiele, B., 2014. A tutorial on human activity recognition using body-worn inertial sensors. ACM Computing Surveys (CSUR), 46(3), p.33.

[8] Zhang X., Yao L., Huang C., Sheng Q. Z., and Wang X.,
“Intent recognition in smart living through deep recurrent neural
networks,” in International Conference on Neural Information
Processing. Springer, 2017, pp. 748–758.

[9] Zhang X., Yao L., Zhang D., Wang X., Sheng Q. Z., Gu T.."Multi-person brain activity recognition via comprehensive EEG signal analysis," in 14th EAI International Conference on Mobile and Ubiquitous Systems: Computing, Networking and Services.

\end{document}